\documentclass[preprint,prb,a4paper,amsmath,amssymb,floatfix,superscriptaddress,showpacs,10pt]{revtex4} 
\usepackage{amssymb}
\usepackage{amsmath}
\usepackage{graphicx}
\usepackage{subfigure}
\usepackage{amsfonts}%
\begin{document}
\title{Magnetic properties of FeCo nanoclusters on Cu(100)}

\author{C.~Etz}
\affiliation{Center for Computational Materials Science,
Technical University Vienna,
A-1060, Gumpendorferstr. 1a, Vienna, Austria}
\author{B.~Lazarovits}
\affiliation{Research Institute for Solid State Physics and Optics
of the Hungarian Academy of Sciences,
Konkoly-Thege M. \'{u}t 29-33., H-1121 Budapest, Hungary}
\author{J.~Zabloudil}
\affiliation{Center for Computational Materials Science,
Technical University Vienna,
A-1060, Gumpendorferstr. 1a, Vienna, Austria}
\author{R.~Hammerling}
\affiliation{Center for Computational Materials Science,
Technical University Vienna,
A-1060, Gumpendorferstr. 1a, Vienna, Austria}
\author{B.~\'Ujfalussy}
\affiliation{Research Institute for Solid State Physics and Optics
of the Hungarian Academy of Sciences,
Konkoly-Thege M. \'{u}t 29-33., H-1121 Budapest, Hungary}
\author{L.~Szunyogh}
\affiliation{Department of Theoretical Physics and
Center for Applied Mathematics and Computational Physics,
Budapest University of Technology and Economics,
Budafoki \'{u}t 8., H-1521 Budapest, Hungary}
\author{G.M.~Stocks}
\affiliation{Materials Science and Technology Division, Oak Ridge National Laboratory, Oak Ridge, Tennessee 37831, USA}
\author{P.~Weinberger}
\affiliation{Center for Computational Materials Science,
Technical University Vienna,
A-1060, Gumpendorferstr. 1a, Vienna, Austria}

\begin{abstract}
We present ab-initio calculations of the magnetic moments and magnetic
anisotropy energies of small FeCo clusters of varying composition on top of a Cu(100) substrate.
Three different cluster layouts have been considered, namely 2$\times$2,
3$\times$3 and cross-like pentamer clusters. The ratio of Co atoms with
respect to the total number in a chosen cluster (``concentration'') was varied
and all possible arrangements of the atomic species were taken into account.
Calculations have been performed fully relativistic using the embedded-cluster
technique in conjunction with the screened Korringa-Kohn-Rostoker method and
the magnetocrystalline anisotropy energy (MAE) has been evaluated by means of the magnetic force theorem.
A central result of the investigations is, that the size of the magnetic moments of
the individual Fe and Co atoms and their contributions to the anisotropy
energy depend on the position they occupy in a particular cluster and on the
type and the number of nearest-neighbors. The MAE for the 2$\times$2 and
3$\times$3 clusters varies with respect to the ``concentration'' of Co atoms in the same
manner as the corresponding monolayer case, whereas the pentamer clusters
show a slightly different behavior.
Furthermore, for the clusters with an easy axis along a direction in the surface plane, 
the MAE shows a significant angular dependence.

\end{abstract}

\maketitle

\section{Introduction}

Surface-supported nanoparticles of magnetic atoms are generating a lot of interest
nowadays, because of potential applications in non-volatile magnetic storage media.
In the last few years, magnetic nanostructures 
were investigated experimentally in terms of various different methods like 
STM, XMCD and MOKE~\cite{Rusponi,Kubetzka,Gambardella,Bergmann,Lau,Repetto}.
In using these techniques together with phenomenological models~\cite{Bruno} and sum-rules~\cite{Carra}
it was possible to infer the high anisotropies and orbital moments
of single magnetic adatoms on a non-magnetic substrate.
Eventually it is the ambition of experimental methods to produce and manipulate nanostructures on an
atom-by-atom level which is
currently attempted by means of for example magnetic tunnel tips.~\cite{Kliewer}
Furthermore, by combining
magnetic and non-magnetic materials like CoPt~\cite{Rusponi}, or two different magnetic species like
FeCo~\cite{Getzlaff}
to form nanoclusters, interesting structural arrangements and enhanced, tunable magnetic properties are obtained.

A key challenge for experimental methods is, in view of the fabrication of so-called
bit-patterned media,~\cite{Richter-1,Brune}
to produce grains of controlled size and position, of known
composition, and with sharply defined magnetic properties.~\cite{Weiss,Held}
This would path the way to increase achievable areal densities~\cite{Coufal}
of magnetic recording media
by several orders of magnitude.~\cite{Mitsuzuka,Richter-2,Sun,Chunsheng,Weiss}

One of the main issues is the magnetic anisotropy energy which
determines the orientation of the magnetization of a cluster with resect to the surface.
Large MAE barriers can stabilize the 
magnetization direction in the cluster and a stable magnetic bit can be created.~\cite{Weller,Bean} 
In contrast to bulk solids, 
clusters deposited on surfaces offer additional degrees of freedom to change the MAE.
It can be influenced by the shape, size and composition of
the cluster and by the substrate. In this paper we will concentrate on the investigation of
the effects of composition and of the details of the atomic arrangements within magnetic clusters
on the anisotropy by means of {\it ab-initio} calculations. 

The information learned from these computational studies can be used
later to construct nanostructures with optimized properties. Moreover,
these studies can be compared to calculations made for
monolayers using the Coherent Potential Approximation (CPA) and
relate the impact of the local environment to the
mean field solution.

\section{Theoretical Approach and Computational Details}

Self consistent, relativistic calculations for FeCo clusters on a Cu(100)
surface have been performed using the embedded-cluster technique~\cite{Bence-1}
within multiple scattering theory (MST) which enables the
treatment of a finite cluster of impurities embedded into a two-dimensional
translational invariant semi-infinite host. Within MST the electronic
structure of a cluster of embedded atoms is described by the so-called
scattering path operator (SPO) matrix given by the following Dyson equation~\cite{Bence-1}
\begin{equation}
\tau_{C}(\epsilon)=\tau_{h}(\epsilon)[1-(t_{h}^{-1}(\epsilon)-t_{C}%
^{-1}(\epsilon))\tau_{h}(\epsilon)]^{-1}\quad ,
\end{equation}
where $\tau_{C}(\epsilon)$ comprises the SPO for all sites of a given finite
cluster $C$ embedded in a host system, $t_{h}(\epsilon)$ and $\tau
_{h}(\epsilon)$ denote the single-site scattering matrix and the SPO of the
unperturbed host sites in cluster $C$, respectively, while $t_{C}(\epsilon)$
stands for the single-site scattering matrix of the impurity atoms. Once
$\tau_{C}(\epsilon)$ is known all corresponding local quantities, i.e., charge
and magnetization densities, spin and orbital moments, as well as the total
energy can be calculated. In all cases the atomic sphere approximation (ASA)
was applied.

Self consistency is achieved by varying the effective potentials and exchange
fields using the local density functionals of Ceperley-Alder
(in the parameterization due to Perdew and Zunger)~\cite{ceperley}
 and solving the Poisson equation as described in Ref.~\onlinecite{Bence-1}.
In all self consistent calculations for the embedded clusters
the orientation of the magnetization was chosen to point uniformly along the
surface normal ($z$ axis). For the calculation of the $t$-matrices and for the
multipole expansion of the charge densities (needed to evaluate the Madelung
potentials), a cutoff for the angular momentum expansion of $l_{\max}=2$ was used.

The host and the cluster sites refer to the positions of an ideal \textit{fcc}
lattice with the experimental lattice constant of Cu ($a=3.6147$ \AA ). 
Structural relaxations of both the cluster--substrate distance as well as of the bond length between
cluster atoms, which may also affect the magnetic properties,~\cite{Pick,Lysenko,Shick,Guirado-Lopez}
have been neglected.
As the atomic radii of the elemental Fe, Co, and Cu are similar, relaxations can
be expected to be relatively small.
For the semi-infinite Cu(100)
host a self consistent fully relativistic calculation was performed in terms
of the screened Korringa-Kohn-Rostoker method~\cite{skkr-1} using 66 $k_{\parallel}$-points
for the irreducible Brillouin zone (IBZ) integration and 16 energy points for
the energy integrations along a semicircular contour in the complex energy
plane by means of a Gaussian quadrature.

The magnetic anisotropy energy (MAE), is defined as the difference in the total energies of the clusters
with the magnetization along two different directions.  While in the bulk or thin films of ferromagnetic
materials the preferred magnetization
direction will be along one of the principle crystal axes, this will not be the case in composite nanoclusters. Depending
on the shape and the composition of the clusters the easy (or hard) axis can be found along a
(arbitrary) direction $\alpha=(\sin{\theta}\cos{\varphi},\sin{\theta}\sin{\varphi},\cos{\theta})$. 
In terms of the magnetic force theorem,~\cite{Jansen} the MAE is given by the energy difference between the band energies
corresponding to two orientations, $\alpha$ and $\beta$, of the magnetization
\begin{equation}
\Delta E_{\alpha\beta}=E_{\alpha}^{b}-E_{\beta}^{b}\quad ,
\end{equation}
which were evaluated using the self consistent potentials taken from the
calculation where the orientation of the magnetization was chosen to be
perpendicular to the surface ($z$ axis).
Here the band energy is actually the grand canonical potential, which at $T=0K$
is given by:%
\begin{equation}
E_{\alpha}^{b}=\int_{\varepsilon_{B}}^{\varepsilon_{F}%
}(\varepsilon-\varepsilon_{_{F}})n(\varepsilon,\alpha)d\varepsilon \quad ,
\end{equation}
where $\varepsilon_{B}$ is the bottom of the valence band and $n(\varepsilon
,\alpha)$ denotes the density of states for the magnetization pointing uniformly along
direction $\alpha$.
Use of the force theorem in the present context seems justified as the atomic species considered are
3d transition metals for which the approximations due to the first order perturbation theory are
legitimate.~\cite{Pastor,Pick,Cabria}

For the present calculations only the nearest-neighbor shell of host and
vacuum sites surrounding the Fe and Co atoms forming the clusters was taken into account.
Our previous convergence studies~\cite{Bence-1} of the magnetic properties of pure Fe
clusters as a function of the number of shells demonstrated, that for a Ag
substrate considering only the first shell suffices. For a Cu host, being a noble
metal with similar properties as Ag,
a first shell of neighbors can therefore safely be assumed to be adequate. In the following,
the term 'cluster' will be used (for simplicity) to denote the set of Co
and/or Fe atoms only.

The dimensions of the clusters were varied in the following way:  we chose a
2$\times$2, a cross-like pentamer and a 3$\times$3 cluster as illustrated, e.g., in Ref.~\onlinecite{Bence-1}.
The atoms in the clusters occupy nearest neighbor positions of the ideal fcc (100) geometry. 
Taking the first shell, the studied systems actually
consist of 30 ASA spheres in the 2$\times$2 case, 40 
for the pentamer cluster and 53 in the case of the 3$\times$3 cluster.
These systems refer to the first 3 layers of a parent $fcc$ lattice. The first
layer is formed by atoms of the substrate, the second layer contains Co and/or
Fe atoms and surrounding vacuum spheres and the third layer is formed only by
vacuum spheres. We varied the 'concentration' of the Co atoms, namely the
ratio of Co atoms with respect to the total number of Co and Fe atoms in the
clusters and considered, in addition, for each concentration also all possible
configurations, i.e. arrangements of the atoms in the cluster.

In the following sections, the most important magnetic properties of the clusters
(detailed for each constituent atom) are discussed, namely the magnetic spin and orbital moments
and the anisotropy energy.
All quantities have been studied with respect to the clusters' size, to the constituent atoms
concentrations and with respect to the different positions of the atoms within
the clusters.
The results are furthermore compared to those of a monolayer of either Co or Fe, in
particular cases also to a monolayer of Co$_{x}$Fe$_{1-x}$ on top of Cu(100).

\section{Results and discussion}

\subsection{Magnetic moments}

One of the central results of all previous studies of (pure) nanoclusters has been that
the spin-only magnetic moments show a negligible dependence on the direction
of the magnetization, while the orbital moments depend significantly on it.~\cite{Bence-1,Bence-2,Stepanyuk,Cabria}
However, it will be shown that in clusters of composite materials, there appears to be a
significant correlation between the size of the spin and the orbital magnetic moments and the specific
positions of the atoms.
As far as the spin moments are concerned, it will be pointed out
that for some clusters this dependency
is also connected with the atomic coordination, and hence it is due to the position occupied by the atom.
In order to present the large amount of data in a compact way, 
below, the results for the magnetic moments are summarized separately for each type of cluster that has been studied.

\subsubsection{2$\times$2 cluster}

\begin{figure}[tbp] \centering
\begin{tabular}{cc}%
\begin{tabular}
[c]{ll}%
{\includegraphics[scale=0.5]
{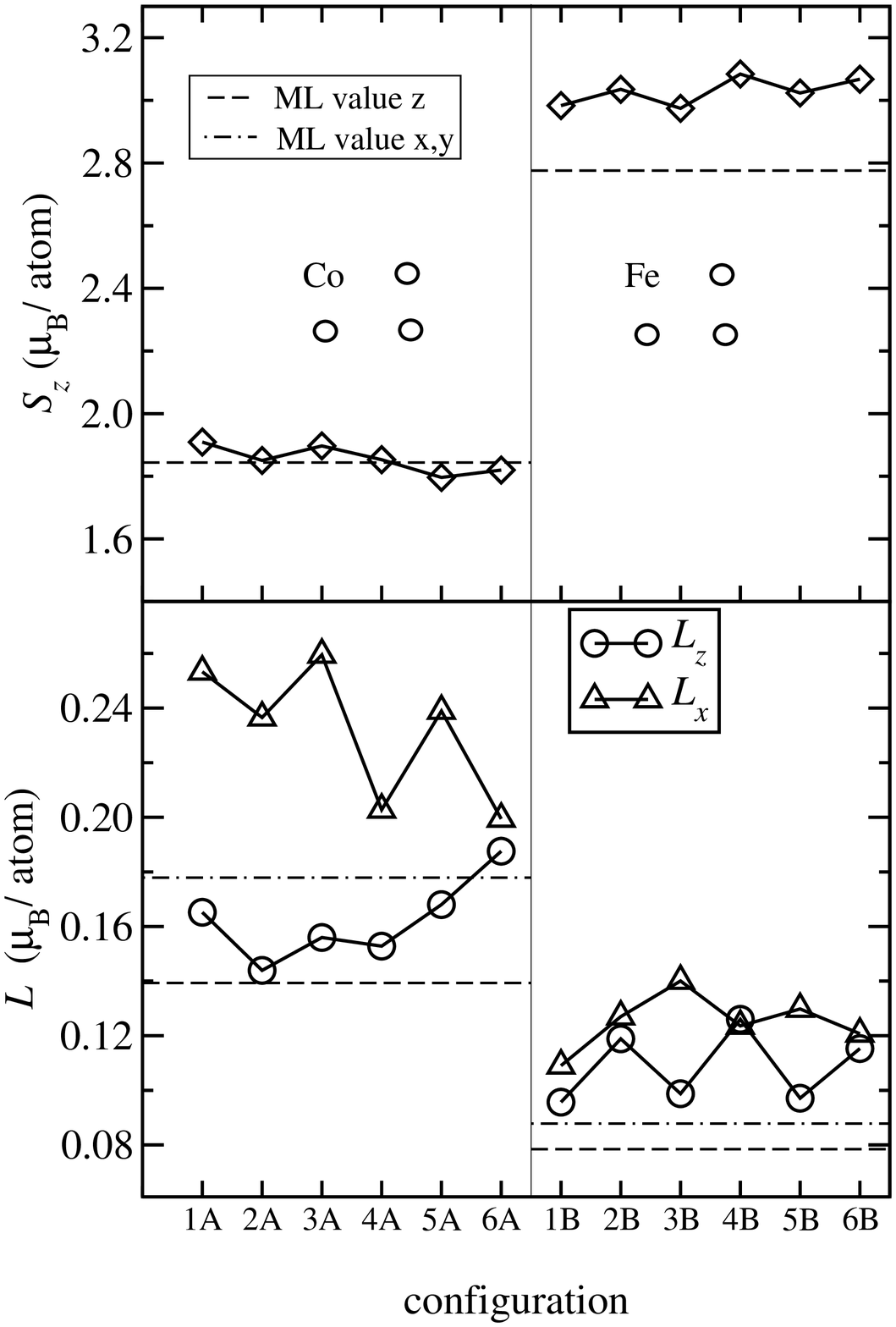}%
}%
&
\raisebox{0.198in}{\includegraphics[scale=0.5]
{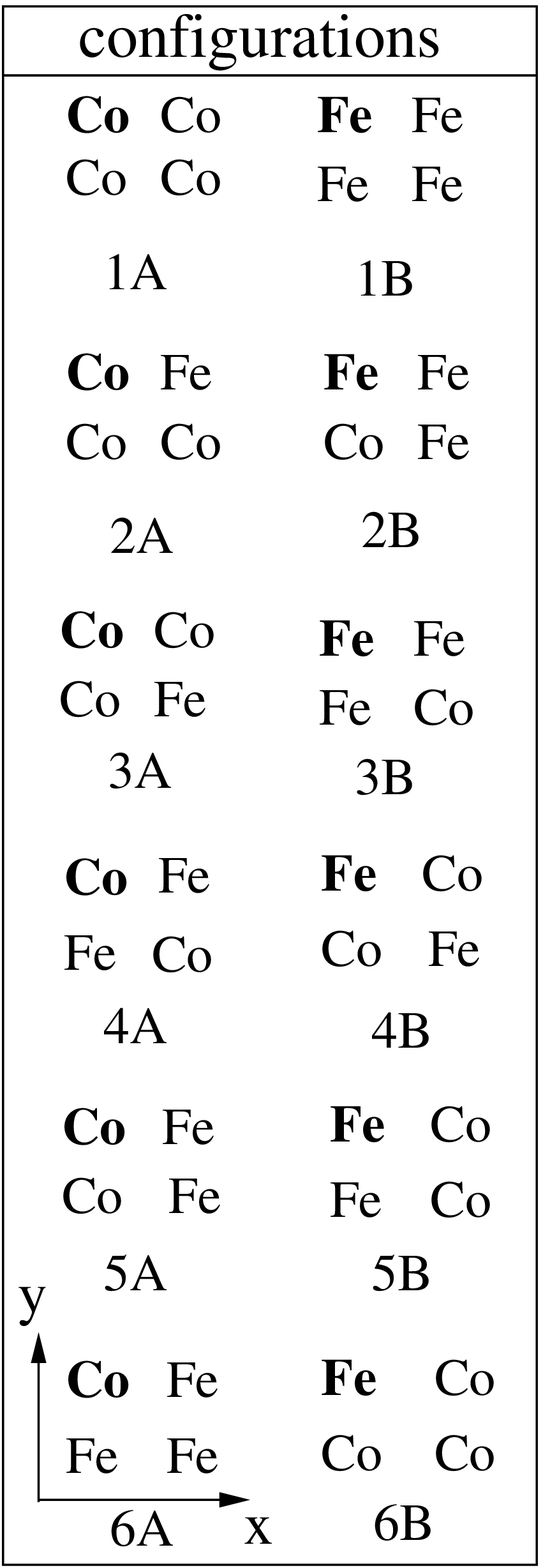}%
}%
\end{tabular}%
\end{tabular}%
\caption{Spin and orbital
magnetic moments of selected Co (left panel) and Fe (right panel) atoms
as a function of all
cluster configurations possible (Co: $i$A, Fe: $i$B, $i=1,\dots,5$), 
see the scheme to the right.
Dashed horizontal lines refer to the corresponding Fe/Cu(100) or Co/Cu(100)
monolayer values for a magnetization along the coordinate \textit{z}-axis.
Dash-dotted horizontal lines to those with the magnetization along the \textit{x}(\textit{y})-axis.
The investigated atom is indicated by boldface letters.}\label{fig1}
\end{figure}
We start our discussion with the magnetic moments for the smallest 
cluster studied, namely for the tetramer. 
In Fig.~\ref{fig1}, the variation of the spin (\textit{S}) and orbital
moments (\textit{L}) of a selected Co and an Fe atom is plotted 
for different cluster configurations (full lines), together with the corresponding Fe or Co
monolayer values (dashed and dash-dotted lines). 
The spin magnetic moment of the chosen Co atom, even in a small 4-atom cluster, 
has almost the same value as in the Co monolayer case, whereas the Fe
magnetic moments in the cluster are enhanced with respect to the Fe 
monolayer due to the reduced coordination number.
In the 2$\times$2 cluster, all the positions are geometrically 
equivalent, however, the orbital magnetic moments are still sensitive 
to the type of the atoms occupying the different positions in the cluster,
or, shortly to the local environment
(see configurations 4A and 5A, 4B and 5B). As can be noticed from the left 
panel of Fig.~\ref{fig1} the orbital moment 
values of the investigated atoms are
more sensitive to the environment than those of the spin moment. 
The value of the orbital moment on an Fe or Co atom can vary by more than 30\% 
depending on its surrounding atoms.
While the spin moments have a very small variation with respect to
the changes in the direction of the magnetization (from the in-plane to
out-of-plane), for the orbital moments the difference can be significant.
For the Co atoms the values of $\textit{L}_x$ are higher than those for $\textit{L}_z$
so it has to be expected that the Co atoms have a preferred in-plane direction for the magnetization. 
The value with respect to $\textit{L}_z$ of the Co atom has a minimum when 
it has one Fe nearest-neighbor (NN) and increases with
the number of Fe atoms in the cluster. Contrary, the same quantity
of an Fe atom shows only minor oscillations when the number of the 
Co NN changes.

\subsubsection{Pentamer cluster}

\begin{figure}[tbp] \centering
\begin{tabular}{cc}%
\begin{tabular}
[c]{ll}%
{\includegraphics[scale=0.5]
{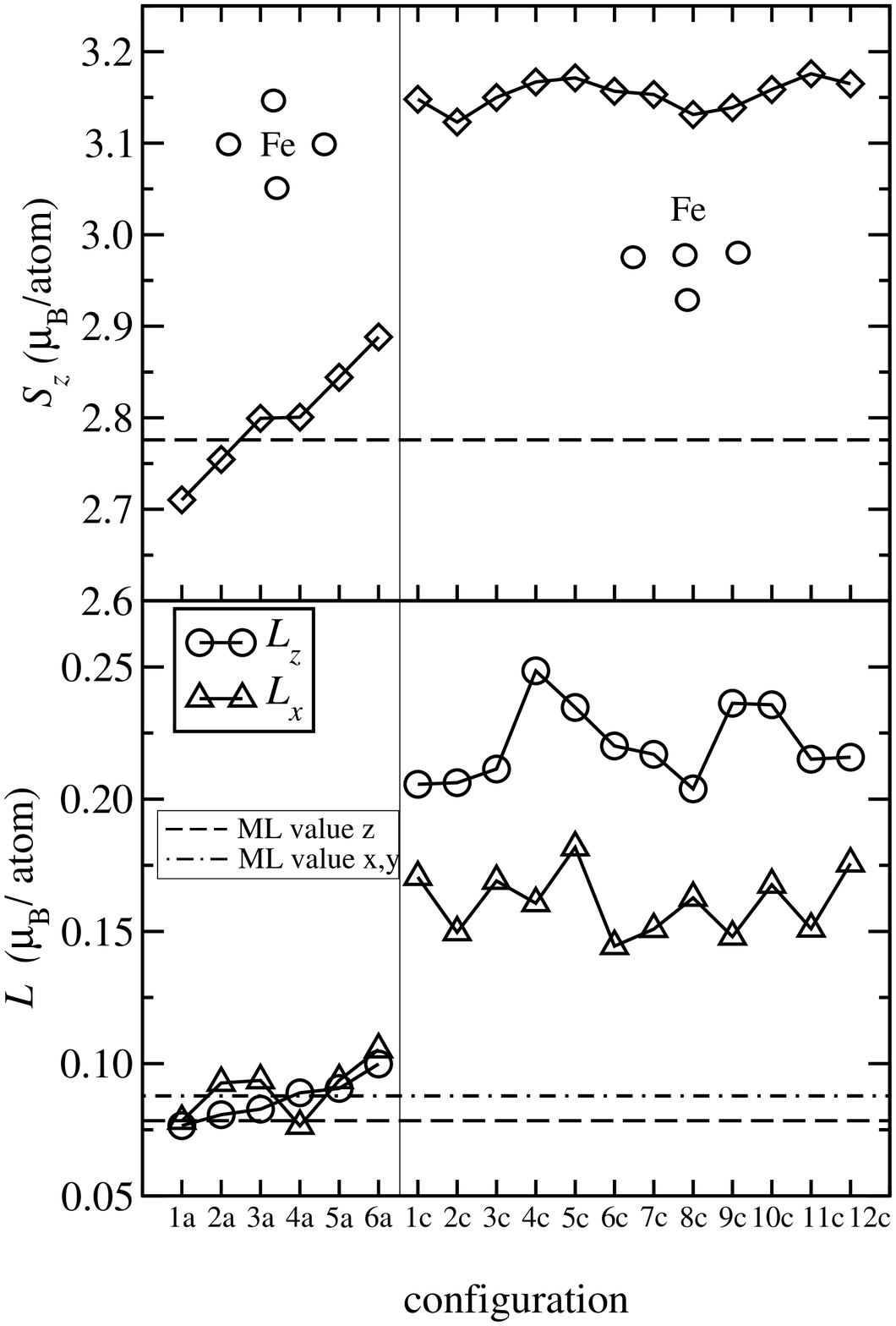}%
}%
&
\raisebox{-0.1773in}{\includegraphics[scale=0.4]
{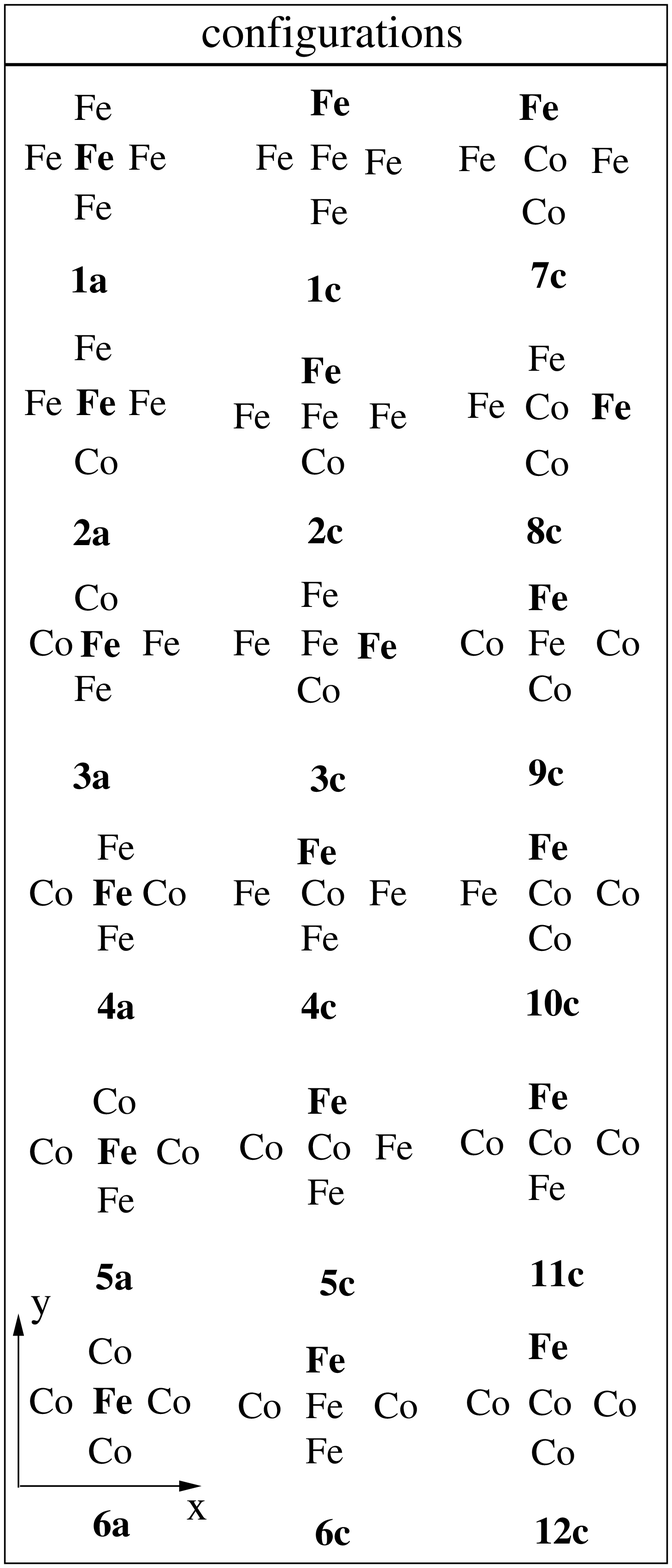}%
}%
\end{tabular}%
\end{tabular}%
\caption{Spin and orbital magnetic moments of a selected Fe atom as a function
of its position in the cluster and of the number of Co neighbors. 
For the actual configuration see the panel to the right. 
Dashed horizontal lines refer to the corresponding Fe/Cu(100) monolayer values with the 
magnetization along the
coordinate z-axis. Dash-dotted horizontal lines to those with the magnetization along the
\textit{x}(\textit{y})-axis.}\label{fig2}%
\end{figure}%

\begin{figure}[ht] \centering
\begin{tabular}{cc}%
\begin{tabular}
[c]{ll}%
\raisebox{-0.0433in}{\includegraphics[scale=0.5]
{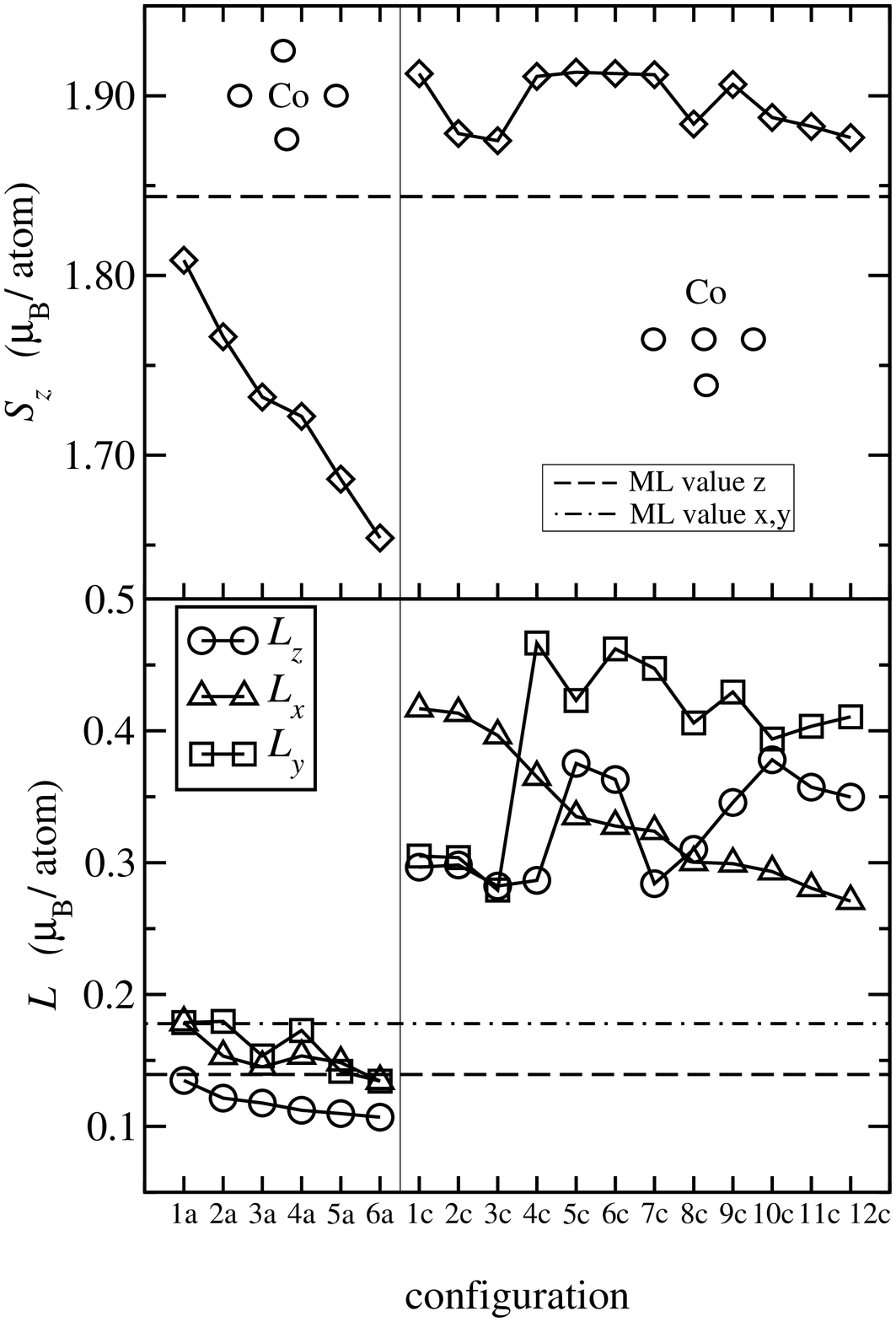}%
}%
&
\raisebox{-0.294in}
{\includegraphics[scale=0.4]
{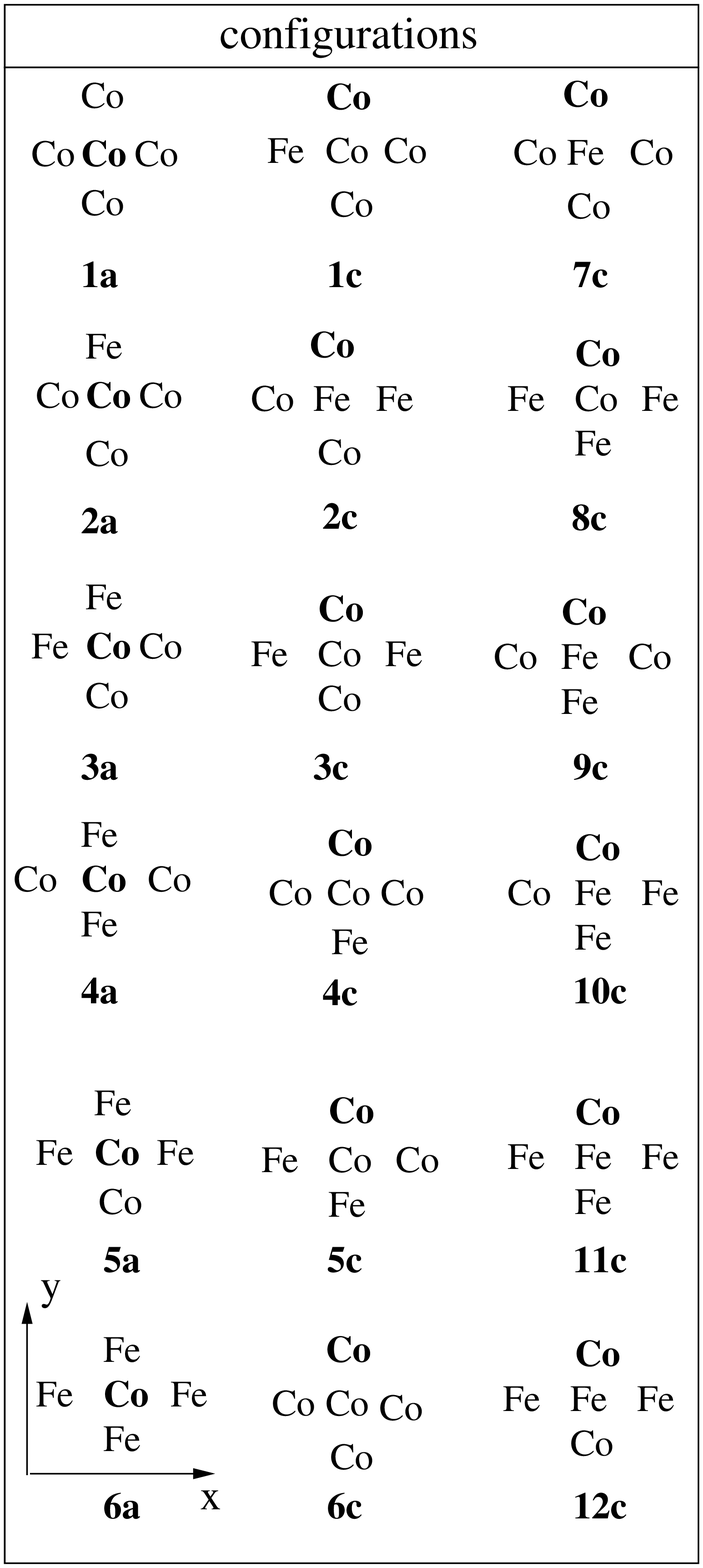}%
}%
\end{tabular}%
\end{tabular}%
\caption{Spin and orbital magnetic moments of a selected Co atom as a function
of the position occupied in the cluster and of the number of Fe neighbors.
For the actual configuration, see the panel to the right.
Dashed horizontal lines refer to the corresponding Co/Cu(100)
monolayer values.
Dash-dotted horizontal lines to those with the magnetization along the x(y)-axis.}\label{fig3}%
\end{figure}%
The distinctive feature of a pentamer cluster - in comparison with a quadratic
2$\times$2 cluster - is the existence of two geometrically nonequivalent 
positions. 
While the atom at the central position in the cluster has four coordinated atoms,
an atom at a corner position has only one.
Fig.~\ref{fig2} illustrates the variation of \textit{S} and \textit{L} of particular Fe atoms with 
respect to changes of neighboring atoms. In the left panel 
the magnetic moments of the central Fe atom are displayed 
and in the right panel those of an Fe atom in a corner position are shown.%

Although the value of the spin and the orbital moments of the central Fe atoms increase
monotonously with the number of Co NNs, they are still in a range of
approximately $\pm0.1\mu_{B}$ with respect to the monolayer value of
$2.85$~$\mu_{B}$ of Fe/Cu(100).
The spin moment shows linear dependence on the number of the Co NNs. 
In contrast, the values of the corner atoms, while being
much higher than in the monolayer case, exhibit a
less pronounced dependence on the environment.%
Taking a closer look at the orbital moments, we see that they are
significantly larger than in the monolayer ($0.078\mu_{B}$) only for
corner atoms for which the value ranges between
$0.2$ and $0.23\mu_{B}$, while
for the central atoms they vary smoothly and are close to the monolayer value.
The difference in the orbital moments values, namely between $\textit{L}_z$ and $\textit{L}_x$ is 
about $0.06\mu_{B}$ and suggests that the corner Fe atoms prefer an out-of-plane 
direction for the magnetization. 
Finally, the spin-moments of an Fe
atom at a corner position is increased dramatically by approximately $0.4\mu_{B}$ as
compared to the value in an Fe monolayer on Cu(100) ($2.77\mu_{B}$).

Fig.~\ref{fig3} shows the same quantities and additionally the orbital moment for
the magnetization pointing long the y-axis, \textit{L}$_y$, now for a Co atom in the central
and at a corner position. Contrary to the case of Fe, the spin-moment
of a central Co atom is smaller than the corresponding monolayer value
($1.84\mu_{B}$) in every configuration. 
Similarly to the Fe case the dependence of \textit{S} 
of the central Co atom on the number of Fe NNs is linear but with opposite sign. 
The decreased spin-moment attains its smallest value when the Co atom
is surrounded by Fe atoms only (configuration 6a in Fig.~\ref{fig3}).
Then again, the magnetic moment of a Co atom positioned at a corner is
larger than in the monolayer. This is not primarily due to the
enhancement of the spin-moment but to that of the orbital moment.
In the lower panel of Fig.~\ref{fig3} the orbital moments are plotted for magnetizations
along the z-direction and along the in-plane x and y directions.
It can be seen, that a central Co atom
shows only a small anisotropy of \textit{L} and a slight dependence on the cluster configuration.
At a corner position, a Co atom exhibits a much more pronounced anisotropy and
for configurations 1c--3c \textit{L}$_x$ is largest,
assuming a given orientation of the clusters with respect to in-plane \textit{x-y} coordinate system
as illustrated in the legend of Fig.~\ref{fig3}.
For all other configurations the orbital moment along the \textit{y}-direction becomes largest and
is significantly enhanced in comparison to the monolayer case.
The reason for the strong increase of the orbital moment of the Fe and Co atoms at corner
positions has been discussed~\cite{Bence-1} for pure Fe, Co, and Ni clusters
and can be traced back to the anisotropic environment and reduced
coordination number. In the present study this effect is 
combined with the changes of the magnetic properties due to the 
different chemical order of the two constituents forming the cluster.

\subsubsection{3$\times$3 cluster}

\begin{figure}[tbp] \centering
\begin{tabular}{cc}%
\begin{tabular}
[c]{cc}%
\raisebox{-0.1565in}{\includegraphics[scale=0.5]
{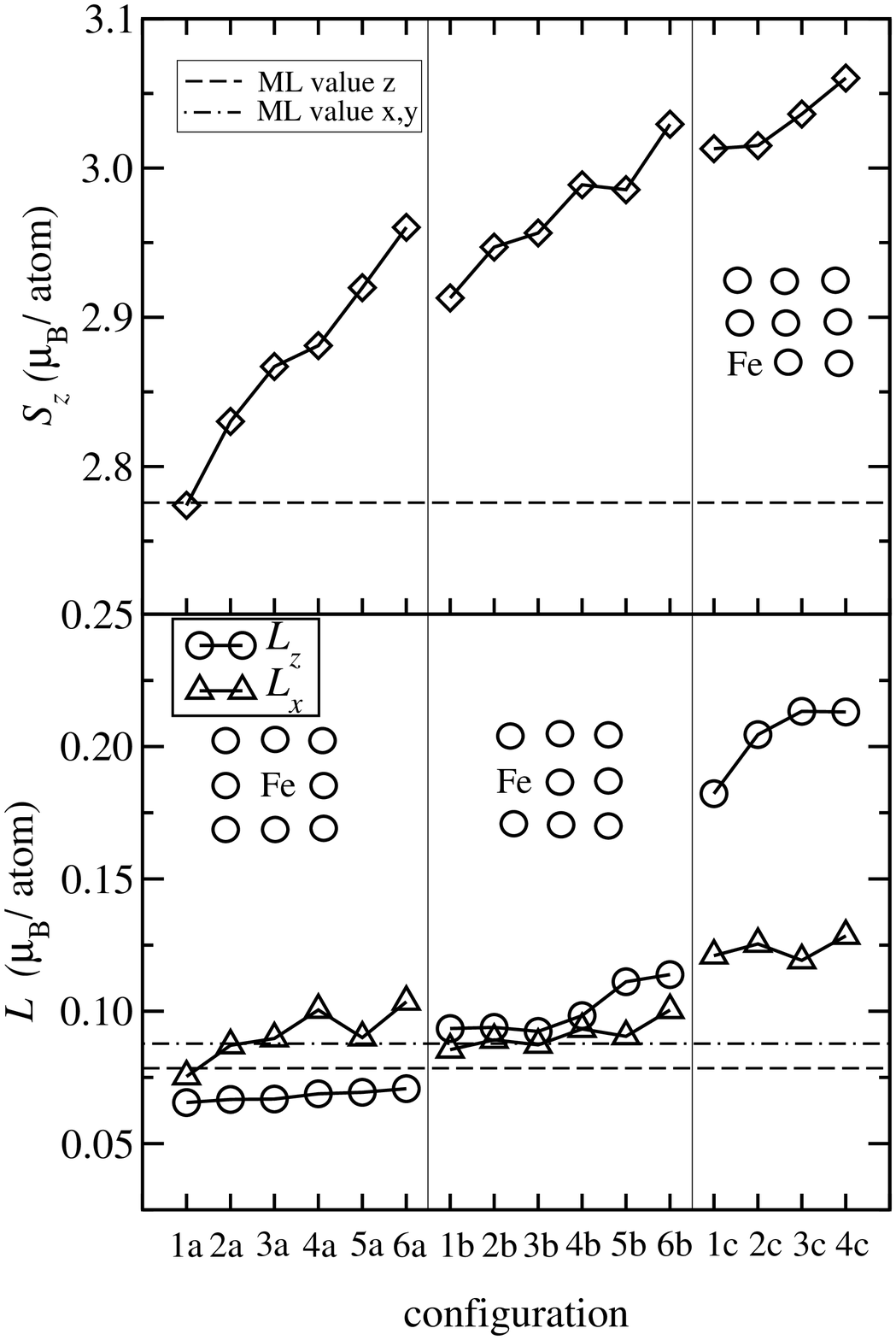}%
}%
&
\raisebox{-0.1706in}{
\includegraphics[scale=0.4]
{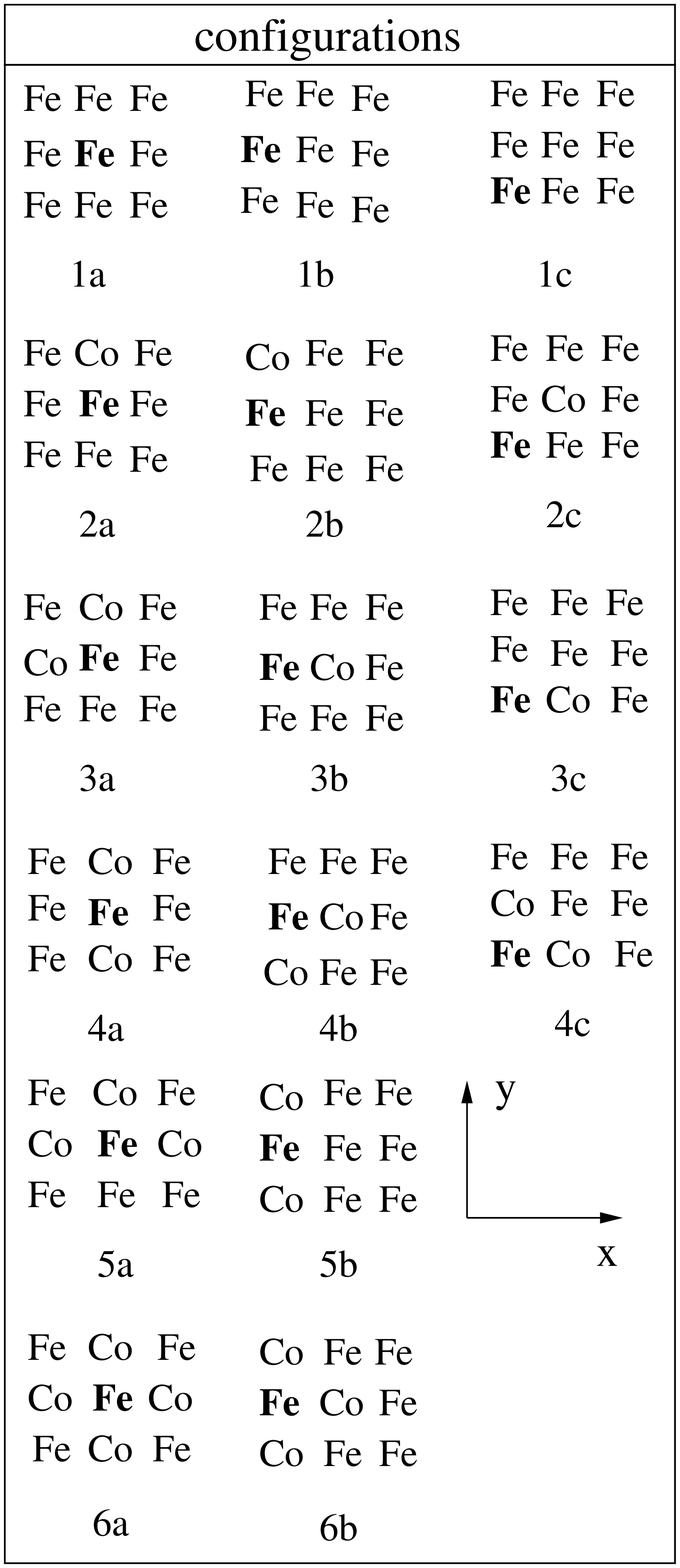}%
}%
\end{tabular}%
\end{tabular}%
\caption{Spin and orbital magnetic moments of a selected Fe atom in a $3\times3$ cluster as a
function of its position and the number of Co NN for selected
cluster configurations (see the panel to the right). Dashed horizontal
lines refer to the corresponding Fe/Cu(100) monolayer values with the
magnetization along the coordinate z-axis.
Dash-dotted horizontal lines to those with the magnetization along the \textit{x}(\textit{y})-axis.}\label{fig4}%
\end{figure}%
\begin{figure}[t] \centering
\begin{tabular}{cc}%
\begin{tabular}
[c]{cc}%
{\includegraphics[scale=0.5]
{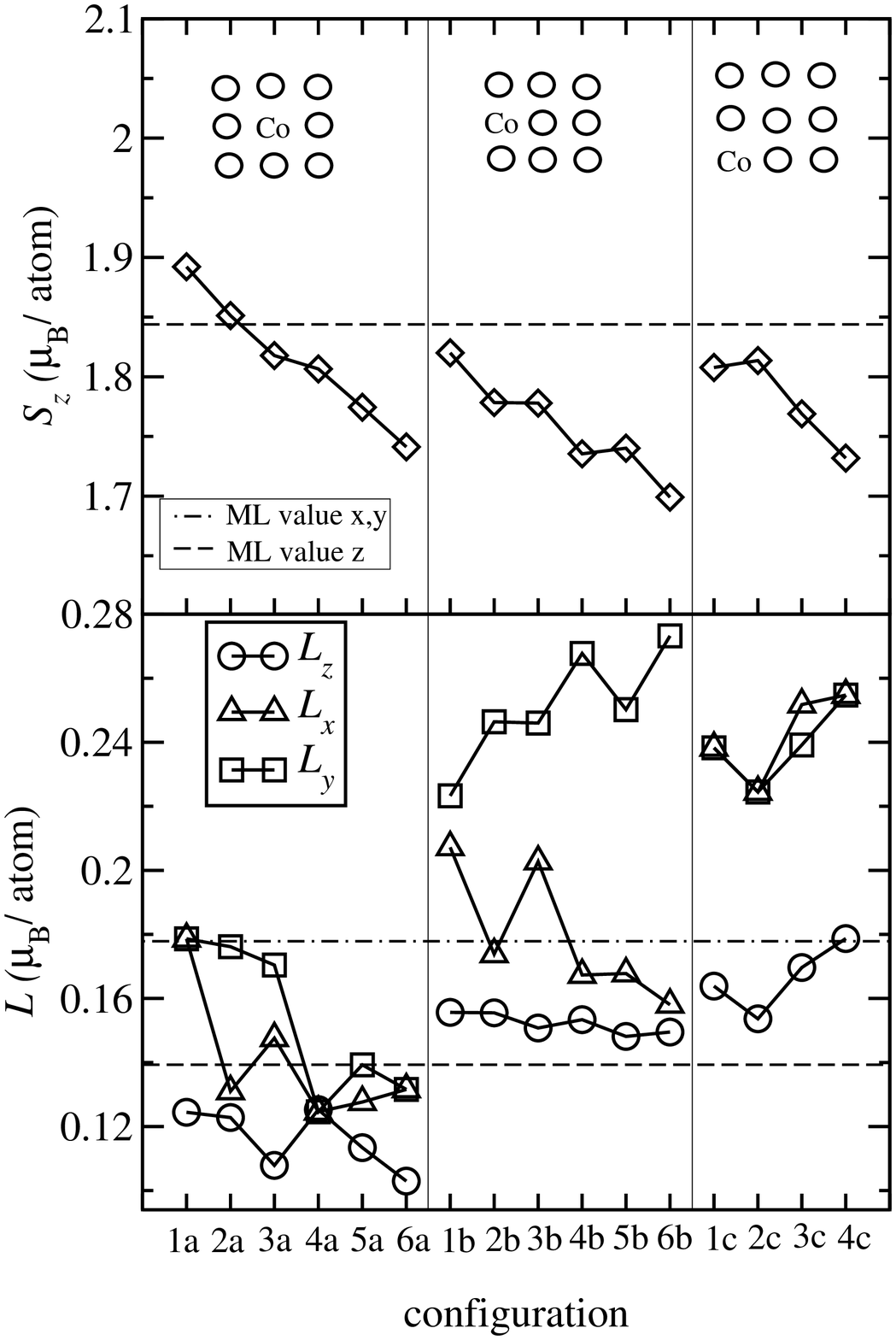}%
}%
&
\raisebox{-0.0518in}{\includegraphics[scale=0.4]
{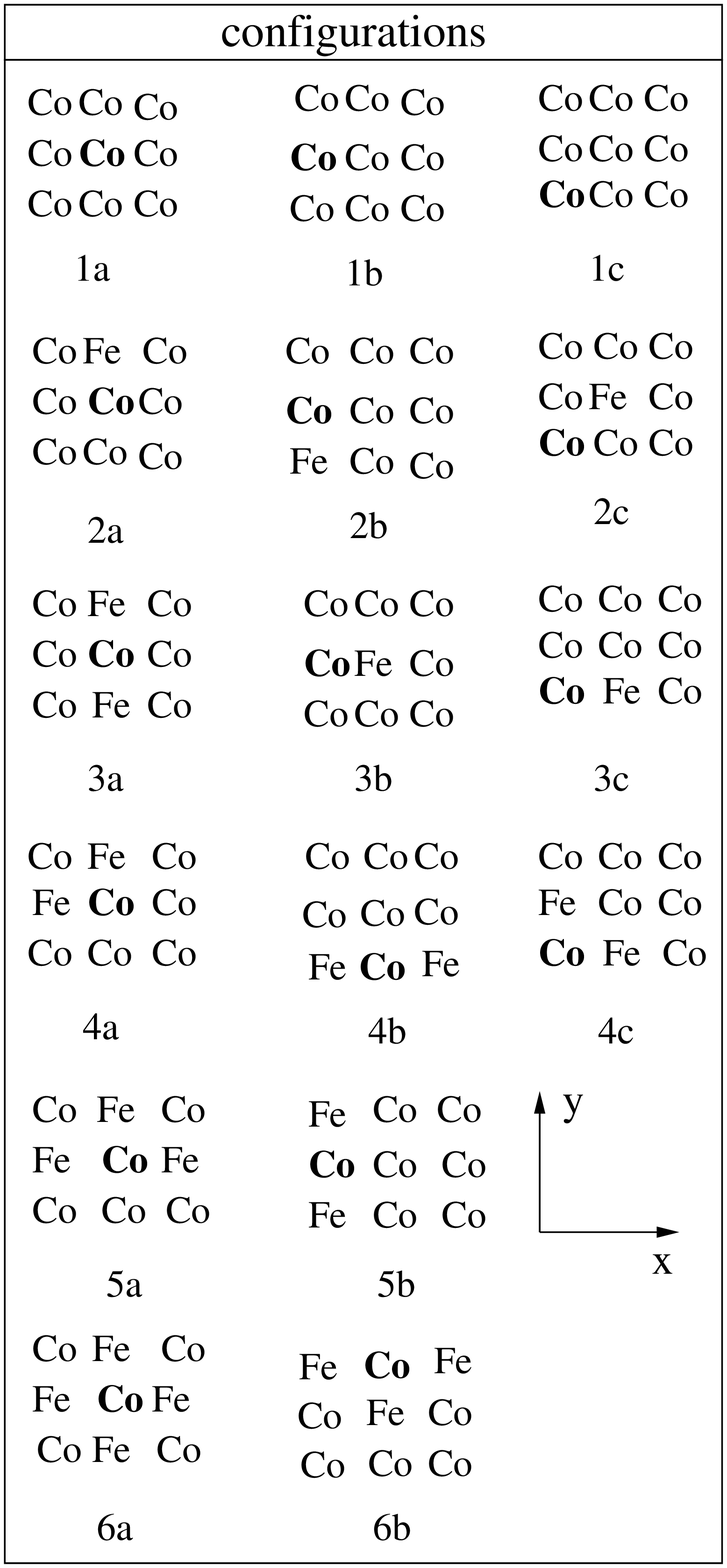}%
}%
\end{tabular}%
\end{tabular}%
\caption{Spin and orbital magnetic moments of a selected Co atom in a $3\times3$ cluster as a
function of its position and the number of Fe NN for selected
cluster configurations (see the panel to the right). Dashed horizontal lines refer to the corresponding
Co/Cu(100) monolayer values with the magnetization along the coordinate z-axis.
Dash-dotted horizontal lines to those with the magnetization along the x(y)-axis.
}\label{fig5}%
\end{figure}%
The magnetic moments for the three geometrically inequivalent positions 
(center, edge, corner) of Fe atoms are displayed in 
Fig.~\ref{fig4} for selected
cluster configurations. Starting in each case from a pure Fe cluster, the NN
positions are successively substituted with Co atoms until all NN sites are
occupied exclusively by Co atoms. 
The only exception is configuration 2c
where a 2nd NN has been considered. 

As a first observation it can be seen, that the spin-moments in the pure clusters
depend on the position of the Fe atom. The values differ by approximately 0.1$\mu_{B}$ when
comparing them at center and center of edge positions.
At a corner position the spin-moment is further enhanced by 0.1$\mu_{B}$ compared to the
edge position. In addition there appears to be an almost linear dependence on the number
of coordinated Co atoms at NN sites.
While it is interesting to observe that the spin-moment of a central Fe atom in a
pure Fe cluster is practically identical to the monolayer value,
the largest moment of 3.06$\mu_{B}$ is found when a Fe corner atom
is coordinated by two Co atoms.

The orbital moments of a central Fe atom are almost identical to their values in a monolayer,
with \textit{L}$_x$ being slightly larger than \textit{L}$_z$, exhibiting negligible dependence
on the local environment.
Similarly, at an edge the orbital moments do not differ significantly from the monolayer values,
and are practically independent of the number and position of Co NN. However, \textit{L}$_z$ is
now augmented in comparison to \textit{L}$_x$.
Most notably the Fe atom located at a corner shows a much more significant enhancement of \textit{L}$_z$
and is further increased as more Co NN are added. 
This can be directly related to the anisotropic
surrounding at that position and reduced coordination number, as was already seen in the case
of smaller clusters.
\textit{L}$_y$ of Fe atoms does not differ significantly from \textit{L}$_x$ and therefore
has not been included in Figs.~\ref{fig1},~\ref{fig2}, and~\ref{fig4}.

Independent of the position assumed by the Co atom in the cluster, the
spin moments decrease almost monotonously with increasing number of Fe NN (see
Fig.~\ref{fig5}). Except for configurations 1a and 2a, all values are smaller than
the monolayer ones.

Again the orbital moments show a dependence on the position that the Co atom occupies in the
cluster. However, \textit{L}$_x$, \textit{L}$_y$, and \textit{L}$_z$ depend in different ways on the addition
of Fe atoms on the NN sites. While at an edge position \textit{L}$_z$ is practically unchanged,
\textit{L}$_x$ shows an oscillatory behavior and \textit{L}$_y$ experiences a strong enhancement as
illustrated in the lower panel of Fig.~\ref{fig5}.
Then again, a Co atom at a corner has a larger \textit{L}$_x$ almost identical to \textit{L}$_y$, 
and the difference to \textit{L}$_z$ is approximately constant.

\begin{figure}[tbp] \centering
\begin{tabular}{cc}%
\begin{tabular}
[c]{l}%
{\includegraphics[scale=0.5]
{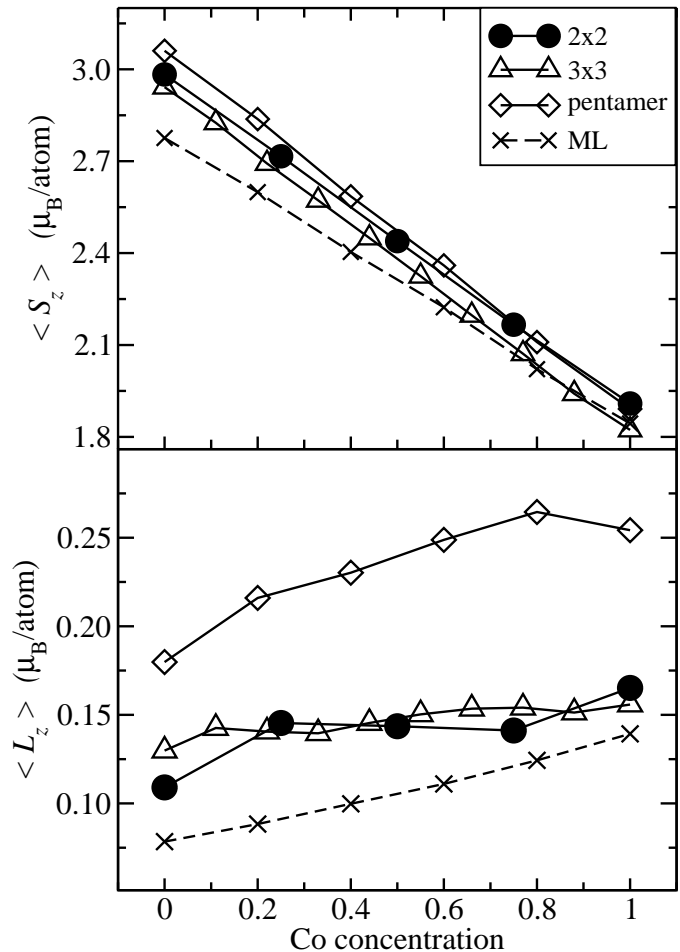}%
}%
\end{tabular}%
\end{tabular}%
\caption{Spin and orbital moments averaged over all the configurations for a given Co
concentration. ($\bullet$: 2$\times$2 cluster, $\Diamond$: pentamer cluster,
$\bigtriangleup$: 3$\times$3 cluster) The dashed line refers to the case of a Co$_x$Fe$_{1-x}$ monolayer on Cu(100).}\label{fig6}%
\end{figure}%
In order to summarize the data discussed above, in Fig.~\ref{fig6} the variation of the 
spin and orbital magnetic moments averaged over all the configurations
for a given Co concentration has been plotted. One can
see that the difference in the averaged values of the 3$\times$3 and 2$\times$2
clusters is very small for both \textit{S}$_z$ and \textit{L}$_z$.
The atoms within a pentamer cluster exhibit a larger
orbital moment with respect to the value of a monolayer of Co$_{x}$Fe$_{1-x}$
on Cu(100), and also compared to the values in the 2$\times$2 and 3$\times$3
cases. 
The variation of the spin moments with respect to the Co concentration
is linear, just as in the monolayer case, whereas the orbital moment has a more complex
behavior. This can be related to the reduced symmetry of the local environment and
to the higher ratio of corner atoms in these clusters.

\subsection{Magnetic anisotropy energy}

\begin{figure}[tb]
\begin{center}
\includegraphics[scale=0.7]
{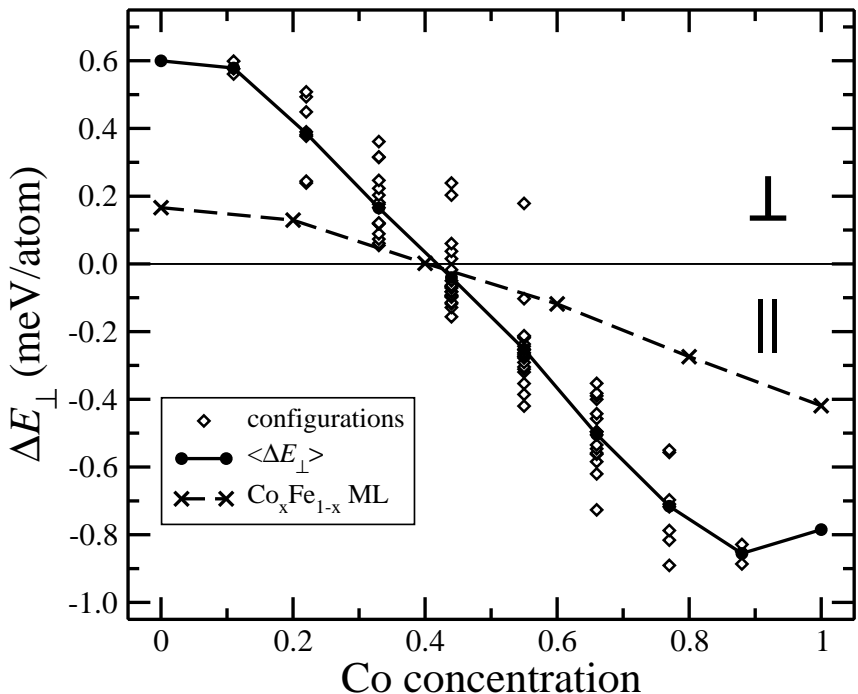}%
\caption{Magnetic anisotropy energy as a function of the Co concentration: 
distribution of the anisotropy energies for 3$\times$3 clusters as a
function of the clusters configuration (diamonds);
the anisotropy energy
$\langle \Delta E_{\perp}\rangle =\langle \left(\Delta E_{xz}+\Delta E_{yz}\right)/2\rangle $, where 
$\langle \rangle $ refers to the average over all configurations (solid line); anisotropy energy for
Co$_{x}$Fe$_{1-x}$/Cu(100) monolayer (dashed line).
The index $\perp$ indicates that the MAE is calculated as the energy difference between 
the (surface) normal \textit{z}
and the in-plane \textit{x} or \textit{y} direction.}%
\label{fig7}%
\end{center}
\end{figure}

\begin{figure}[tb]
\begin{center}
\includegraphics[scale=0.6]%
{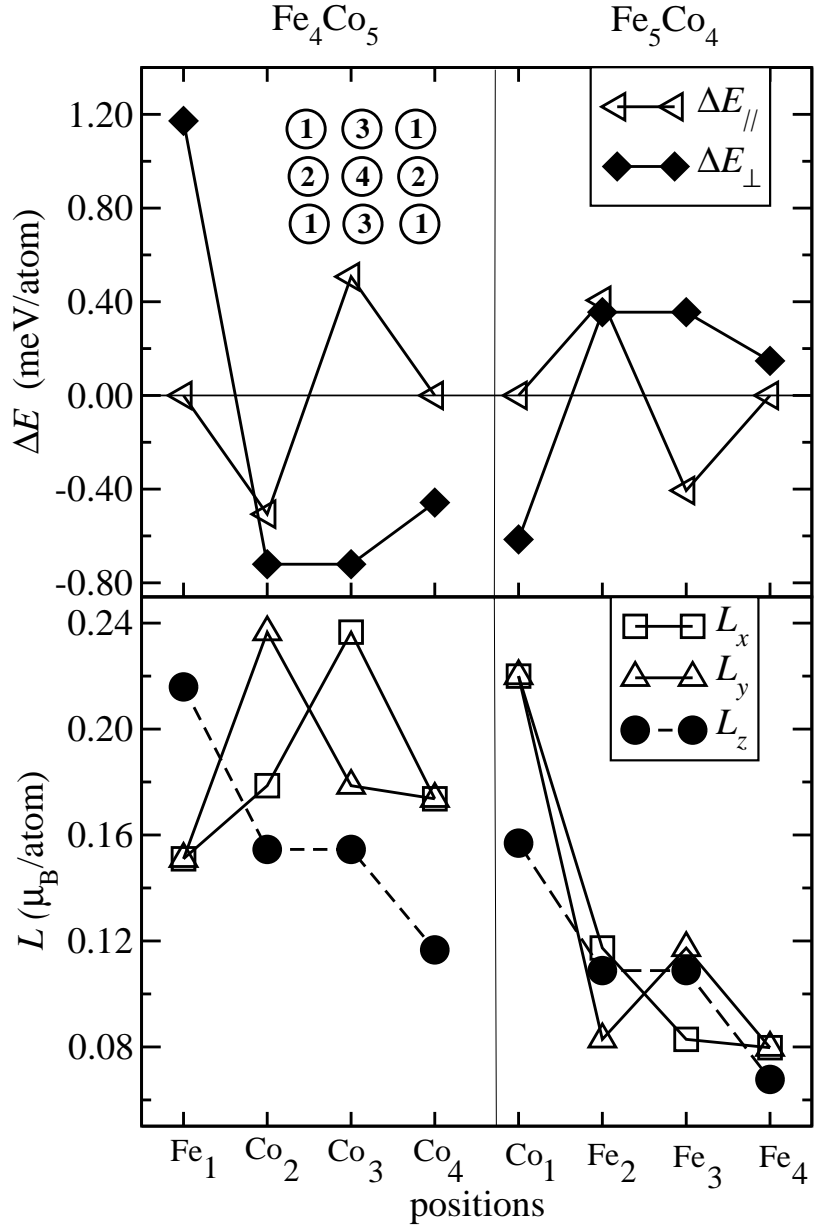}%
\caption{MAE contributions (upper panel) and orbital magnetic moment variation (lower panel) per atom for
Fe$_{4}$Co$_{5}$ and Fe$_{5}$Co$_{4}$. Full circles, squares and triangles refer in turn to
$\textit{L}_z$, $\textit{L}_x$ and $\textit{L}_y$.
In the case of
Fe$_{4}$Co$_{5}$, the Fe is in position 1 (denoted by Fe$_1$) and Co are in positions: 2, 3 and 4
(denoted in turn by Co$_1$, Co$_2$, and Co$_3$);
in the case of Fe$_{5}$Co$_{4}$, the Co is in position 1 and the Fe are in
positions 2, 3 and 4. ($L^{Co}$ in Co/Cu(100) is 0.139$\mu_{B}$/atom;
$L^{Fe}$ in Fe/Cu(100) is 0.08$\mu_{B}$/atom). Triangles and full
diamonds refer to $\Delta$E$_{\parallel}$ and $\Delta$E$_{\perp}$, respectively.
$\Delta E_{\parallel}=\Delta E_{yx}$ and 
$\Delta E_{\perp}=\left(\Delta E_{xz}+\Delta E_{yz}\right)/2$.
The indices $\perp$ and $\parallel$ indicate that the MAE is 
calculated once for the magnetization switch from the normal \textit{z}
towards the in-plane \textit{x} (\textit{y}, respectively) direction and
then for the in-plane anisotropy between the \textit{x} and the \textit{y} direction.}%
\label{fig8}%
\end{center}
\end{figure}

\begin{figure}[tb]
\begin{center}
\includegraphics[scale=0.6]
{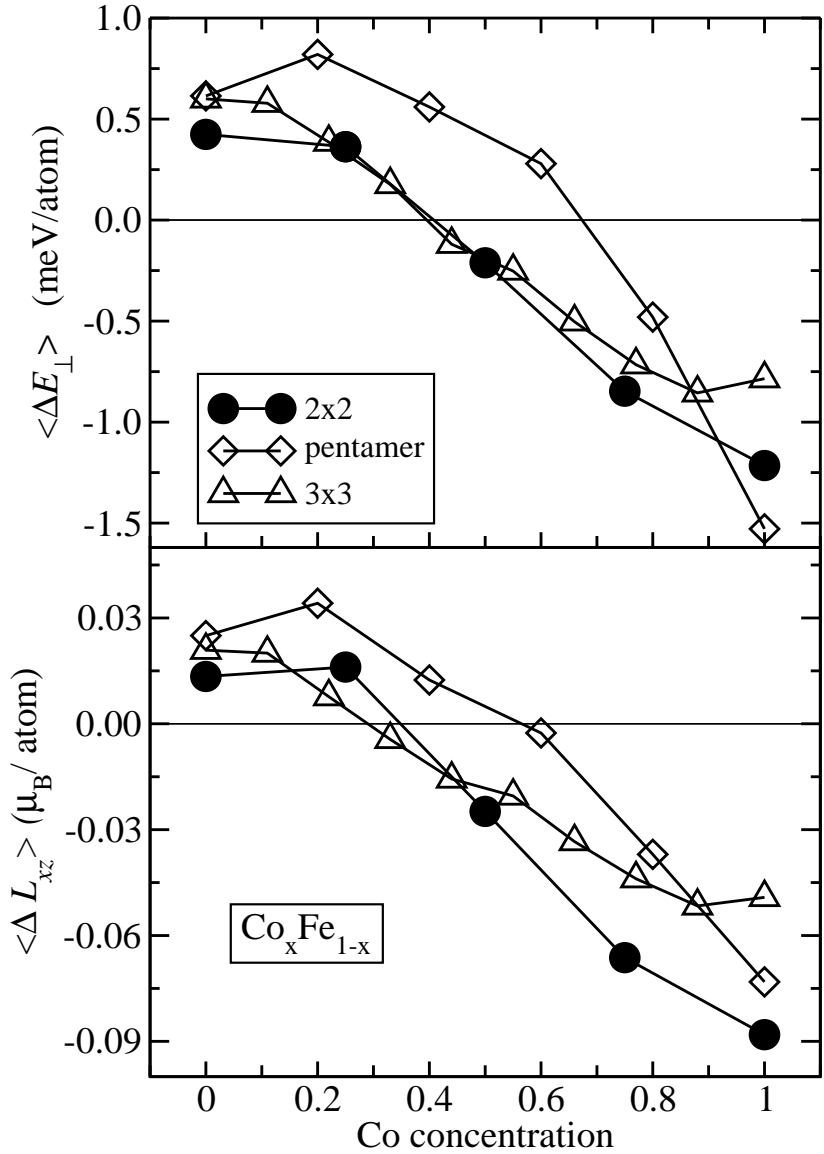}%
\caption{Upper panel: MAE per atom for the 2$\times$2, 3$\times$3 and the pentamer
cluster, averaged over all configurations with respect to the Co
concentration.
$\langle \Delta E_{\perp}\rangle =\langle \left(\Delta E_{xz}+\Delta E_{yz}\right)/2\rangle $ corresponds to the
average over all configurations at a given concentration. The index $\perp$ 
shows that the MAE is calculated with respect to the magnetization along the surface
normal \textit{z} and 
the in-plane \textit{x} or \textit{y} direction. Lower panel:
anisotropy of the averaged orbital magnetic moment for a 2$\times$2, a
pentamer and a 3$\times$3 cluster. $\langle \Delta L_{xz}\rangle =\langle L_{x}-L_{z}\rangle $, where $\langle \rangle $ refers to 
the average over all configurations. }%
\label{fig9}%
\end{center}
\end{figure}

\begin{figure}[tb]
\begin{center}
\includegraphics[scale=0.7]
{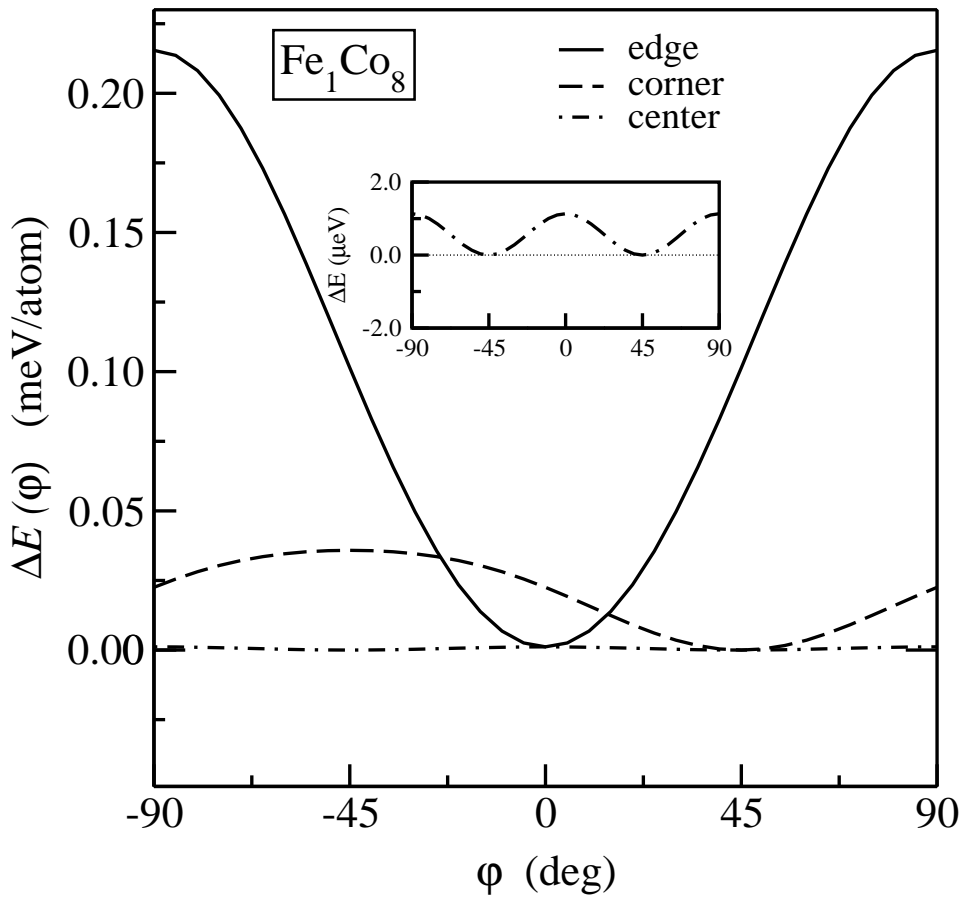}%
\caption{Dependence of the anisotropy energy, $\Delta E(\varphi)=E(\varphi)-E(\varphi_0)$,
with respect to different directions of the magnetization,
specified by the azimuth angle $\varphi$.
The three curves stand for the different configurations (c.f. Fig.~\ref{fig12}):
no 1, corner Fe atom (dashed line); 
no 2, central Fe atom (dash-dotted line), no 3, edge Fe atom (solid line)}%
\label{fig10}%
\end{center}
\end{figure}

\begin{figure}[tb]
\begin{center}
\includegraphics[scale=0.7]
{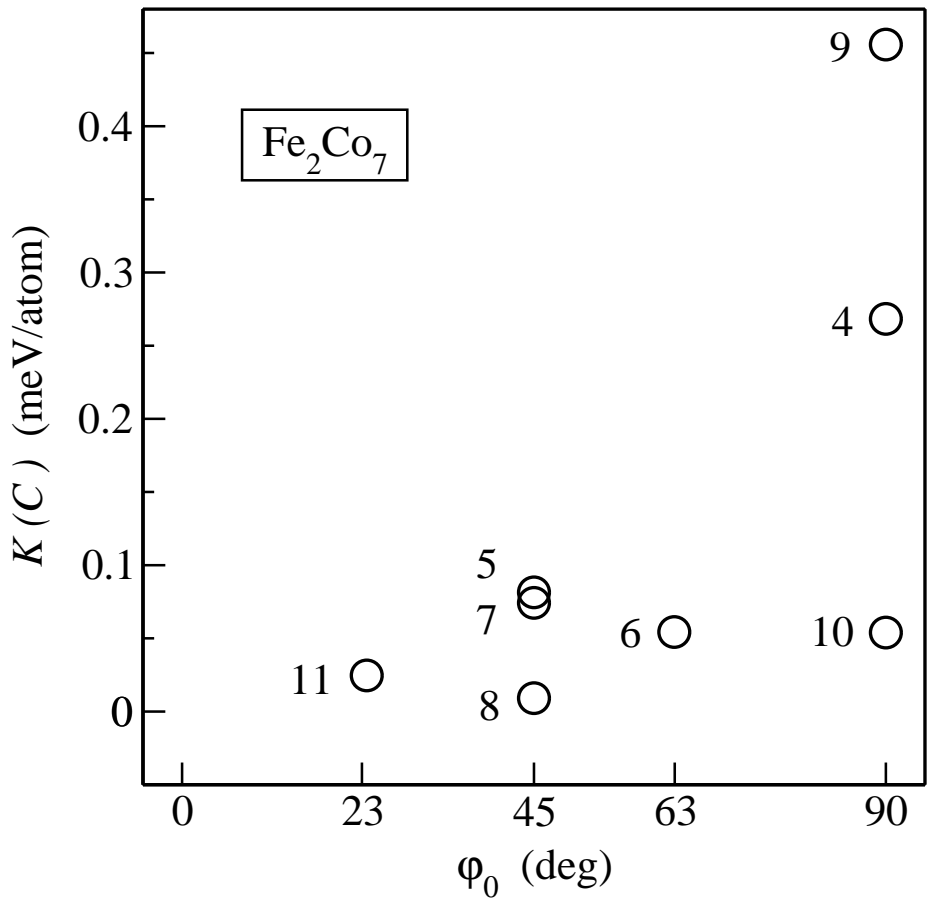}%
\caption{The anisotropy constants $K(\mathcal{C})$
for eight different configurations of Fe$_2$Co$_7$ cluster.
The specified angles $\varphi$$_0$ refer to the direction of the easy axis of magnetization 
with respect to the cluster configurations. For each case, 
the corresponding configuration number (c.f. Fig.~\ref{fig12}) is indicated next to the symbols.}%
\label{fig11}%
\end{center}
\end{figure}

\begin{figure}[tb]
\begin{center}
\fbox{\includegraphics[scale=0.6]
{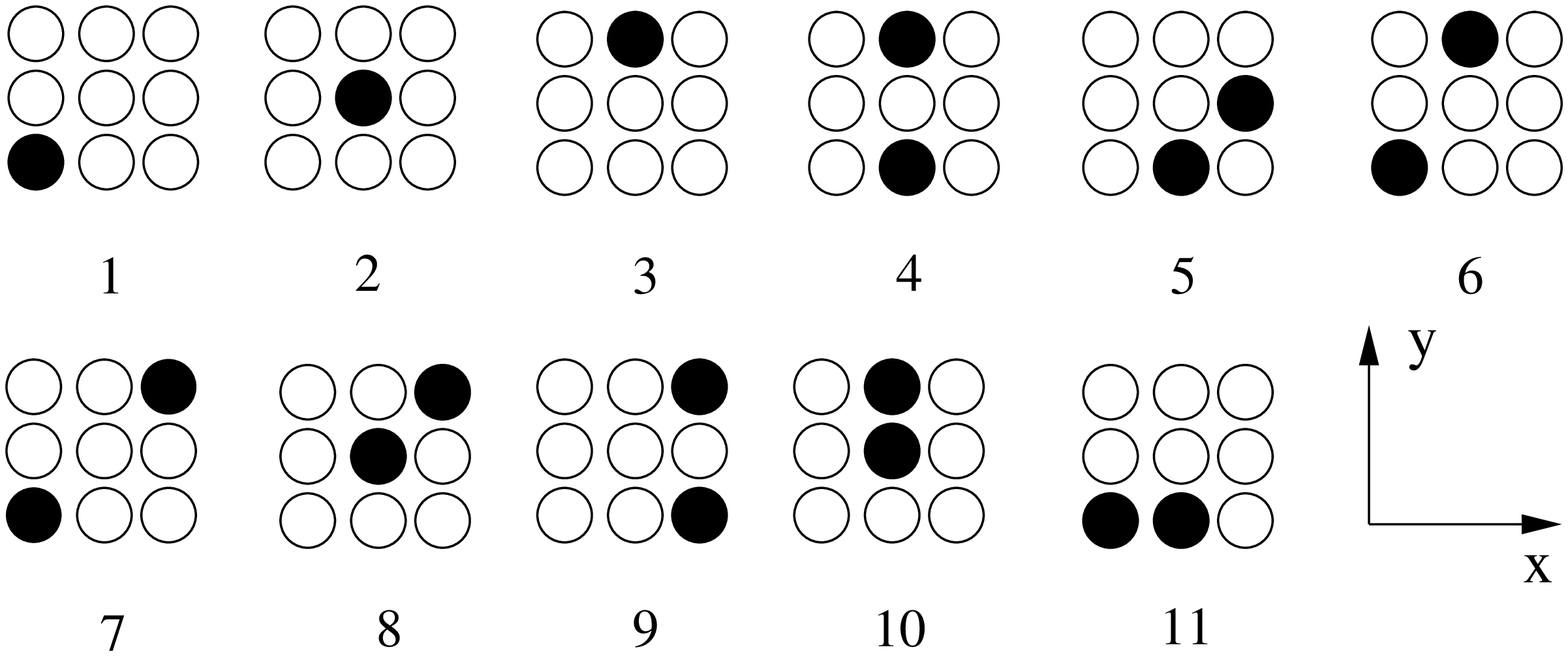}}%
\caption{Inequivalent configurations of 3$\times$3 clusters containing one or two atoms, respectivley,
of a foreign species.  Filled (empty) circles represent Fe (Co) atoms and vice versa.} 
\label{fig12}%
\end{center}
\end{figure}
The magnetic anisotropy arises from that part of the total energy which is 
anisotropic with respect to two different magnetization directions.~\cite{Brooks} One contribution 
comes from the shape anisotropy (also referred to as demagnetization or dipole-dipole 
energy) which is, however, quite small (in the $\mu$eV range) if the clusters are small and of
square shape.
Due to the strong anisotropy of the orbital moments, and the dependence of both, the spin and
the orbital moments on the local 
environment, this energy is nevertheless enhanced in comparison to pure clusters.
However, it is negligible compared to the most important contribution to
the magnetic anisotropy in this case, namely the magnetocrystalline anisotropy
energy which is caused by spin-orbit coupling of the electrons in a magnetic material.
This energy is rather sensitive to the local environment and due to the arrangement of atoms, 
a magnetization along certain orientations is energetically preferred. Subsequently it will be
shown, that the magnetocrystalline anisotropy is closely related to the structure and symmetry
of the investigated clusters.

The MAE of the three cluster types has been investigated with respect to
perpendicular orientations of the  magnetization along the coordinate axes,
as well as according to the dependence on the azimuth angle
$\varphi$.
To proceed systematically, our discussion will start with the magnetization along \textit{x},\textit{y}, 
and \textit{z}, and will then 
focus on the angular dependence of the MAE.

We compare the results for the 3$\times$3 clusters with those for one monolayer,
in Fig.~\ref{fig7}, where the configuration averaged MAE with respect to in-plane
(along the \textit{x}-axis) and out-of-plane (along the \textit{z}-axis) magnetization, 
$\langle \Delta E\rangle _\perp=\langle E_{x}^{b}-E_{z}^{b}\rangle $, is displayed  
as a function of the Co concentration. 
The size of the anisotropy energy is strongly influenced by specific cluster configurations.
In some cases the type of the configuration may even be responsible for a change 
in the preferred orientation of the easy axis
as is the case of $x=0.44\%$ Co 
and $x=0.55\%$ Co.
Starting from a pure Fe cluster, which has a large anisotropy energy (+0.6 meV/atom),
favouring out-of-plane orientation, the increase in the Co concentration is responsible for a decreasing anisotropy 
energy until the easy axis prefers an in-plane orientation. 
Then a further increase in the Co concentration leads to 
an augmentation of the in-plane anisotropy. 
The general trend, however, is very similar to the one obtained in a CPA calculation for a monolayer of Co$_{x}$Fe$_{1-x}$/Cu(100). 
In comparison to the MAE calculated for a monolayer of Co$_x$Fe$_{(1-x)}$/Cu(100) (dashed line) 
the values for the 3$\times$3 clusters are about three times larger.

It has been suggested by Bruno~\cite{Bruno} and Gambardella~\cite{Gambardella,Gambardella-1}
that the direction of
the easy axis is strongly related to the magnitude of the orbital moments.
We investigated the orientation of the MAE with respect to the variation of the orbital moments for
the 3$\times$3 clusters, by choosing a configuration with high planar symmetry
and applying it to two different concentrations (Fe$_{4}$Co$_{5}$, Fe$_{5}$Co$_{4}$).
In order to investigate the correspondence between the orbital moment anisotropy and
the MAE, in Fig.~\ref{fig8}, the values of the MAE and of the orbital moment per
atom are plotted with respect to the positions occupied in the cluster.
For the two selected clusters, 
three different magnetization directions (\textbf{M}%
$\parallel$$z$, \textbf{M}$\parallel$$x$, \textbf{M}$\parallel$$y$ ) were considered.
The rule is strictly followed in the case of Fe$_{4}$Co$_{5}$ and stands for almost all
atoms in Fe$_{5}$Co$_{4}$, except for the central Fe (i.e., Fe$_{4}$) which
actually has a very small orbital moment anisotropy (Fig.~\ref{fig8}, upper panel).
The equivalent positions within a 3$\times$3 cluster (labeled by the
same number) are schematically represented in the upper part of
Fig.~\ref{fig8}. For sake of simplicity, in this figure only the values
for nonequivalent positions are presented. The nonequivalence between positions 2 and 3 
is only due to the planar symmetry, which can also be seen from 
the interchange of the values for \textit{L}$_x$ and \textit{L}$_y$. 
While the Fe atoms give an out-of-plane contribution to the MAE, the Co atoms have a tendency to an
in-plane magnetization. 
The values of the atomic contributions (from both Fe and Co), vary with the position of the atom within the cluster
and increases with increasing coordination number. This example illustrates that the preferred magnetization direction of the cluster as 
a whole results from
an interplay of contending tendencies of the cluster constituents.

It is interesting to compare also the configuration averaged MAE,
which is thought to be measured in experimental situations 
of the 2$\times$2 and 3$\times$3 with those of the pentamer clusters as in Fig.~\ref{fig9}. One
notices that the MAE of the 2$\times$2 and 3$\times$3 clusters have the same
tendency as in the monolayer case, whereas for the pentamer clusters the
situation is different. The larger values for most concentrations in that case 
are due to the high ratio of corner atoms which have a reduced coordination
and whose contributions to the MAE are consequently the largest.
In all the cases we have a change of direction for the easy axis 
with respect to the concentration.
For the 2$\times$2 and 3$\times$3 clusters this change occurs at approximately
the same concentration ($\sim$ 40\% Co), while for the pentamer cluster, the
Co concentration at which the MAE direction changes is higher ($\sim$ 70\%).
In the lower panel of Fig.~\ref{fig9} we show that the behavior of the MAE follows closely the
averaged orbital moment anisotropy.

The previous sections have focused on the energy difference between the directions 
along the coordinate axes \textit{x}, \textit{y} and \textit{z} only. However, 
one can expect uniaxial anisotropies of the in-plane magnetization along 
intermediate directions due to the symmetry of a specific cluster configuration.
Hence in the following we investigate the angular dependence -- specified 
by the azimuth angle $\varphi$ -- of the MAE for selected clusters, in order
to illustrate the complex behavior of the in-plane anisotropy.

Fig.~\ref{fig10} shows the dependence of the total MAE on the azimuth angle for
the simple case of one Fe atom in a $3\times3$ Co cluster. 
Within this cluster the Fe atom can occupy three inequivalent sites -- corner, edge, center -- and 
the direction of the easy axis, as well as the MAE strongly depends on the position of the Fe atom.
For the cluster with the higher symmetry with respect to the \textit{x} and \textit{y} axis (Fe atom at center), 
the value of the anisotropy is very close to zero and its angular variation is only due to
the cluster symmetry, which has a periodicity of 90$^\circ$ (inset of Fig.~\ref{fig10}).
If the Fe atom is located at the center of an edge the anisotropy energy is two orders of magnitude higher.
In case the Fe atom sits on an edge parallel to the \textit{x}-axis, the easy magnetization axis is also
along the \textit{x} direction and the hard axis is perpendicular to it along the \textit{y} direction.
The periodicity of the angular 
variation of the MAE in this case is $180^{\circ}$, and the energy needed to switch the
magnetization between two easy axis is identical to the amplitude.
If the Fe atom is positioned at a corner, then the easy axis is along $\varphi$=$45^{\circ}$.
Consequently the angular dependence of the MAE can be approximated by:
$\Delta E(\varphi)=K(\mathcal{C})\sin^{2}(\varphi-\varphi_{0})$
, where $\varphi_{0}$ is the direction of the easy magnetization axis, depending on the configuration
symmetry, and $K(\mathcal{C})$ is an anisotropy constant, which is very sensitive to the specific
cluster configuration $\mathcal{C}$.

As an additional illustration, Fig.~\ref{fig11} shows the values of the anisotropy
constants $K(\mathcal{C})$ and the angles
$\varphi_{0}$ of the easy axis for Fe$_2$Co$_7$. This cluster is distinguished from the previously 
discussed one by the appearance of two additional angles at $\varphi_{0}\approx$23$^{\circ}$
and $\varphi_{0}\approx$63$^{\circ}$ corresponding to symmetry axes of the clusters.
The number at the left of each symbol refers to the corresponding cluster configurations as given 
in Fig.~\ref{fig12}.

These simple cases illustrate the general features of 
more complicated ones, namely that the direction of the easy axis is determined from the symmetry
of the cluster configuration, and that the size of the anisotropy energy depends
on the specific distribution of atoms.
Clearly this means, that by averaging over a large collection of clusters, where the exact individual
configurations are not known, it is neither possible to predict the anisotropy energy, nor the direction of
the easy axis of a single nanocluster.

\section{Conclusion}

In this work we have determined the magnetic moments and the
magnetic anisotropy of three different types of composite FeCo
nanoclusters on a Cu(100) substrate by means of \textit{ab-initio} calculations.
The study has concentrated on the effect of composition on the cluster properties, neglecting
structural relaxation effects.
While the atomic potentials are approximated as being spherically
symmetric, higher moments of the charge density have been taken into account.  
Clusters of three different geometries have been considered -- 2$\times$2, cross-like pentamer,
and 3$\times$3 -- and the positions of the Fe and Co atoms have been
varied according to all possible arrangements for given ``concentrations''.

As in clusters of a pure material, the size of the spin magnetic moments depends on the position in 
the cluster for both, Fe and Co atoms, and consequently on the number of coordinated atoms as has
been noticed previously by Mavropoulos~\cite{Mavropoulos}.
In addition, depending on the atoms' position and on the geometry
of the cluster, the spin-moments show either no, or a high sensitivity to changes in the local environment,
i.e., with respect to changes of the number or the position of the foreign atomic species.

Orbital moments are strongly influenced by the position of an atom. 
As has been found in many previous 
calculations~\cite{Bence-1,Bence-2,Pick-1,Dorantes-Davila,Nonas,Reddy,Wildberger,Minar,Bornemann,Cabria,Stepanyuk,Xie}
low coordinated
atoms exhibit a significant increase of their orbital moments.  In the present study
it could be shown, that they also depend on the specific cluster configurations and 
on the concentration of the atomic species.
Moreover, low coordinated Co atoms, which have the largest moments within the surface plane,
also exhibit a strong in-plane anisotropy of their orbital moments.
Perimeter Fe atoms, in contrast, have strongly enhanced orbital moments along the z-direction, perpendicular
to the substrate surface.

By averaging over all possible cluster configurations, the total magnetic moments show a variation
with respect to the concentration of Co atoms, which is very close to that found in monolayers of
Fe$_x$Co$_{1-x}$ as obtained from CPA calculations.

When evaluating the magnetic anisotropy energy it appears that the orientation of the easy axis
depends on the cluster symmetry, originating from the distribution of the two different magnetic atoms.
Then again, the size of the anisotropy energy is highly sensitive to the specific arrangements of
atoms. Clusters which have a preferred in-plane magnetization direction are found to have
an unusually large angle dependent anisotropy energy. In these cases the direction of the
easy axis will not necessarily be along the crystal axes, but is determined by the specific
distributions of Fe and Co atoms.

By plotting the MAE, averaged over all configurations, as a function of concentration of one atomic species,
it is found that the easy magnetization direction of a collection of clusters can be tuned
in a similar manner, as in thin films of the same materials,~\cite{Zab98}
by varying the concentration of the constituents.

\section{Acknowledgment}

Financial support to this work was provided by the Austrian Science Foundation (WK W004),
the Center for Computational Materials Science (Contract No. Zl. 98.366),
the Hungarian National Scientific Research Foundation (OTKA NF061726 and T046267),
and ORNL (subcontract No.~4000055828).

\end{document}